\documentclass[onecolumn]{svjour3}
\smartqed

\usepackage{fix-cm}
\usepackage{graphicx}
\usepackage{lipsum,amsmath}
\usepackage{flushend}

\begin{document}

\title{Dynamical evolution of quintessence cosmology in a physical phase space}

\author{Jing-Zhao Qi  \and  Ming-Jian Zhang \and  Wen-Biao Liu }
\institute{ Jing-Zhao Qi \at Department of Physics, Institute of Theoretical Physics, Beijing Normal University, Beijing, 100875, China\\
           \and
           Ming-Jian Zhang \at Department of Physics, Institute of Theoretical Physics, Beijing Normal University, Beijing, 100875, China\\
           \and
           Wen-Biao Liu \at Department of Physics, Institute of Theoretical Physics, Beijing Normal University, Beijing, 100875, China
  \email{wbliu@bnu.edu.cn}
}

\date{Received: date / Accepted: date}
\maketitle

\begin{abstract}
The phase space analysis of cosmological parameters $\Omega_{\phi}$ and $\gamma_{\phi}$ is given. Based on this, the well-known quintessence cosmology is studied with an exponential potential $V(\phi)=V_{0}\exp(-\lambda\phi)$. Given observational data, the current state of universe could be pinpointed in the phase diagrams, thus making the diagrams more informative. The scaling solution of quintessence usually is not supposed to give the cosmic accelerating expansion, but we prove it could educe the transient acceleration. We also find that the differential equations of system used widely in study of scalar field are incomplete, and then a numerical method is used to figure out the range of application.
\end{abstract}

\keywords{cosmology  \and  quintessence  \and cosmic expansion history}

\section{Introduction}  \label{introduction}
The late time cosmic acceleration has been supported by many independent cosmological observations, including the type Ia supernovae (SNIa) \cite{riess1998supernova}, large scale structure \cite{tegmark2004cosmological}, cosmic microwave background (CMB) anisotropy \cite{spergel2003wmap} etc. Dark energy is believed to have unveiled the mystery of the cosmic acceleration, via the ratio of its pressure and energy density as well as the equation of state (EoS) parameter $w<-1/3$. Depending whether EoS is constant or time-dependent, the dark energy candidate can be divided into many subclasses. The cosmological constant model ($\Lambda$CDM) with $w=-1$ is the most robust model but with the fine-tuning problem \cite{weinberg1989cosmological,weinberg2000cosmological} and coincidence problem \cite{1999PhRvL..82..896Z}. This led to a widespread speculation that the EoS of vacuum energy is not a constant. The quintessence model \cite{Caldwell:2005tm,Zlatev:1998tr,Tsujikawa:2013fta} was naturally introduced to study the possible existence of one or more scalar fields.

The well-known quintessence scenario is described by a canonical scalar field $\phi$ minimally coupled to gravity. The quintessence slowly evolving down a potential $V(\phi)$ can lead to the cosmic acceleration, which is similar to slow roll inflation in the early universe. The tracker solution of quintessence makes the universe to be insensitive to the initial conditions, thus alleviating the coincidence problem \cite{1999PhRvL..82..896Z}. The dynamics of quintessence will have different characteristics under different potentials, and they have been studied in detail \cite{1998PhRvD..57.4686C,2000PhRvD..61l3503D,2001PhRvD..64h3510N,2014PhRvD..89h3521T}. The phase space analysis is a most common and effective method, and has been used to investigate dynamical behavior of cosmological model, such as quintessence \cite{1998PhRvD..57.4686C,2014arXiv1412.5701L}, tachyons \cite{2012EPJC...72.2095Y,2004PhRvD..69l3502A}, \textit{k}-essence \cite{2011CQGra..28f5012Y}, quintom \cite{2012arXiv1208.0061L}, phantom \cite{2005JCAP...05..002G}, modified gravity \cite{2013PhRvD..87h4031T,2012PhLB..712..430W}, and so on. In these literatures, the authors analyzed the dynamical evolution of a universe by the global attractor. However, their analyses are presented using some intermediate variables, such as $x-y$, whose definitions have no explicit physical significance; as a consequence, it is hard to discern the evolution of the universe.

A transformation form $x-y$ to $\Omega_{\phi}-\gamma_{\phi}$ is more explicit in physical significance, which also be used widely in study of scalar field. And then we try to analyse the evolution of cosmology in $\Omega_{\phi}-\gamma_{\phi}$ space by employing this transformation. This cosmological parameters' space shows the evolutionary history of the quintessence cosmology in a phase diagram with more explicit physical significance. Moreover, given observational data, more information can be obtained from the diagram. However the appearance of singularity indicates that the transformation is incomplete. Therefore, it needs to determine the scope of application for this transformation.

This paper is organized as follows. In Sec. \ref{quintessence}, the quintessence cosmology and the exponential potential are reviewed, and both forms of differential equations describing autonomous system are listed as well. In Sec. \ref{comparison}, two analytical methods based on two sets of differential equations are compared. In Sec. \ref{singularity}, the singularity problem is described and discussed in detail. Finally, in Sec. \ref{end},  discussion and conclusion will be given.

\begin{table*}[htb]
\begin{center}
\begin{tabular}{|c|c|c|c|c|c|c|c|}
\hline
Point & $x$ & $y$ & Existence & $w_{\rm eff}$ & Acceleration & $\Omega_\phi$ & Stability \\
\hline
\hline
 $O $ &  0 &  0 &  $\forall\;\lambda,w_{b} $ &  $w_{b} $ &  No &  0 &  Saddle \\ & & & & & & & \\
\hline
 $A_- $ &  -1 &  0 &  $\forall\;\lambda,w_{b} $ &  1 &  No &  1 & Unstable if $\lambda\geq-\sqrt{6}$ \\ & & & & & & & Saddle if $\lambda<-\sqrt{6}$ \\
\hline
 $A_+ $ &  1 &  0 &  $\forall\;\lambda,w_{b} $ &  1 &  No &  1 &  Unstable if $\lambda\leq \sqrt{6}$\\ & & & & & & & Saddle if $\lambda>\sqrt{6}$ \\
\hline
$B $ & $\sqrt{\frac{3}{2}}\frac{1+w_{b}}{\lambda} $ & $\sqrt{\frac{3(1-w^2_{b})}{2\lambda^2}} $ & $\lambda^2\geq 3(1+w_{b}) $ & $w_{b} $ & No & $\frac{3(1+w_{b})}{\lambda^2} $ & Stable \\ & & & & & & & \\
\hline
$C $ & $\frac{\lambda}{\sqrt{6}} $ & $\sqrt{1-\frac{\lambda^2}{6}} $ & $\lambda^2<6 $ & $\frac{\lambda^2}{3}-1 $ & $\lambda^2<2$ & 1 & Stable if $\lambda^2<3(1+w_{b})$ \\ & & & & & & & Saddle if $3(1+w_{b})\leq\lambda^2<6$ \\
\hline
\end{tabular}
\end{center}
\caption[crit]{\label{crit} The properties of the critical points (This table is taken from Ref. \cite{2014PhRvD..89h3521T}).}
\end{table*}

\section{The quintessence cosmology} \label{quintessence}

Under the assumption of a flat and homogeneous universe, the Friedmann-Robertson-Walker (FRW) metric in units of $8\pi G=c=1$ could be written as
\begin{eqnarray}
ds^2=-dt^2+a^2\left[dr^2+r^2(d\theta^2+\sin^2\theta d\phi^2)\right].
\end{eqnarray}
We consider a quintessence universe in the presence of a background prefect fluid which is described by EoS $w_{b}=p_{b}/\rho_{b}$ ranging from $0$ (dust) to $1/3$ (radiation), where $p_{b}$, $\rho_{b}$ respectively represents the pressure and energy density of background fluid. The Friedmann equations of quintessence cosmology are
\begin{eqnarray}
H^{2}&=&\frac{1}{3}(\rho_b+\rho_{\phi}),\label{Fried1}\\
\dot{H}&=&-\frac{1}{2}\left(\rho_{b}+\rho_{\phi}+p_{b}+p_{\phi}\right)\label{Fried2},
\end{eqnarray}
where $H$ is the Hubble parameter, $\rho_{\phi}$ is the energy density of scalar field, a dot represents the derivative with respect to cosmic time $t$.
The pressure and energy density of quintessence are given by
\begin{equation}
p_{\phi}=\frac{\dot{\phi}^2}{2}-V(\phi)
\end{equation}
and
\begin{equation}
\rho_{\phi}=\frac{\dot{\phi}^2}{2}+V(\phi).
\end{equation}

The corresponding continuity equation of scalar field is
\begin{equation}\label{ceq}
\ddot{\phi}+3H\dot{\phi}+\frac{\partial V}{\partial \phi}=0.
\end{equation}
The EoS parameter of quintessence is
\begin{equation}
w_{\phi} \equiv \frac{p_{\phi}}{\rho_{\phi}}=\frac{\frac{\dot{\phi}^2}{2}-V(\phi)}{\frac{\dot{\phi}^2}{2}+V(\phi)}.
\end{equation}

For $V \gg \dot{\phi}^2$ the quintessence approaches to cosmological constant model because $w_{\phi}\sim -1$, while for $V \ll \dot{\phi}^2$ the quintessence describes a stiff fluid with $w_{\phi}\sim 1$.

In order to transform the cosmological equations into an autonomous dynamical system, the most frequently used method is to introduce some auxiliary variables  \cite{2001PhRvD..64h3510N}
\begin{equation}
x=\frac{\dot{\phi}}{\sqrt{6}H}, ~~~~~~ y=\sqrt{\frac{V}{3H^2}}, ~~~~~~ \lambda=-\frac{1}{V}\frac{dV}{d\phi}.
\end{equation}
So Eqs. (\ref{Fried1}), (\ref{Fried2}) and (\ref{ceq}) can be rewritten as
\begin{eqnarray}
\label{x}
x'&=&\sqrt{\frac{3}{2}}\lambda  y^2+\frac{3}{2}x(x^2-y^2-1) \nonumber\\
&&+\frac{3}{2}w_b x(1- x^2-y^2), \\
\label{y}
y'&=&-\sqrt{\frac{3}{2}}\lambda xy+\frac{3}{2}y(1+x^2-y^2) \nonumber\\
&&+\frac{3}{2}w_b y(1-x^2-y^2), \\
\label{l}
\lambda'&=&-\sqrt{6}\lambda^2(\Gamma-1)x,
\end{eqnarray}
where
\begin{equation}
\label{Gamma}
\Gamma \equiv V \frac{d^2 V}{d\phi^2}/\left(\frac{dV}{d\phi} \right)^2,
\end{equation}
and the prime denotes the derivative with respect to $\ln a$.

Introducing the scalar field energy density $\Omega_{\phi}$ and the effective equation of state $\gamma_{\phi}=1+w_{\phi}$ as the following expressions
\begin{eqnarray}
\Omega_{\phi}&=&\frac{\rho_{\phi}}{3H^2}= x^2+y^2, \\
\gamma_{\phi}&=&\frac{\rho_{\phi}+p_{\phi}}{\rho_{\phi}}=\frac{\dot{\phi}^2}{V+\dot{\phi}^2/2}=\frac{2x^2}{x^2+y^2},
\end{eqnarray}
Eqs. (\ref{x})-(\ref{l}) could be modified as \cite{2014PhLB..731..342G}
\begin{eqnarray}
\label{omega}
\Omega'_\phi&=&3(\gamma_b-\gamma_\phi)\Omega_\phi(1-\Omega_\phi),\\
\label{gamma}
\gamma'_\phi&=&(2-\gamma_\phi)(-3\gamma_\phi+\lambda\sqrt{3\gamma_\phi\Omega_\phi}),\\
\label{lambda}
\lambda'&=&-\sqrt{3\gamma_\phi\Omega_\phi}\lambda^2 (\Gamma-1).
\end{eqnarray}
Here, $\gamma_b=1+w_b$ is the effective EoS of background prefect fluid. The above system (\ref{omega})-(\ref{lambda}) becomes an autonomous system while the tracker parameter $\Gamma$ is a function of the roll parameter $\lambda$, which is widely used to study scalar field \cite{scherrer2008thawing,gupta2014thawing,chiba2013observational,2014PhLB..731..342G}. But it should be noted that the above equations are tenable only in $x\geq 0$. If $x<0$, $\sqrt{3\gamma_\phi\Omega_\phi}$ should be $-\sqrt{3\gamma_\phi\Omega_\phi}$. In general, people use these equations just on the case $x\geq0$.

The value of $\Gamma$  depends on the form of potential $V(\phi)$, and it could be a constant when the potential $V(\phi)$ takes a special form, such as $\Gamma=0, 0.5, 1$ for $V(\phi)=V_{0}\phi, V_{0}\phi^2, V_{0}\exp(-\lambda\phi)$, respectively. Here, $V_0$ is a constant. In this paper, the phase space analysis of cosmological parameters $\Omega_{\phi}$ and $\gamma_{\phi}$ is given. Based on this, the quintessence cosmology is studied with an exponential potential $V(\phi)=V_{0}\exp(-\lambda\phi)$. Under this potential, the parameter $\lambda$ is a constant and the above mentioned 3-dimensional autonomous system becomes 2-dimensional as a result.

With regard to this exponential potential, the dynamical evolution was analyzed by employing dynamical system techniques \cite{1998PhRvD..57.4686C,2014PhRvD..89h3521T}. The critical points of the system and their properties are displayed in Table \ref{crit}. Here $w_{eff}$ is the effective EoS parameter, which is defined as
\begin{eqnarray}
w_{eff}&\equiv&\frac{p_{\phi}+p_b}{\rho_{\phi}+\rho_b} \nonumber \\
&=&\frac{p_{\phi}}{\rho_{\phi}+\rho_b}+\frac{p_b}{\rho_b}\frac{\rho_b}{\rho_{\phi}+\rho_b} \nonumber \\
&=&x^2-y^2+w_b(1-x^2-y^2). \label{weff}
\end{eqnarray}
If $w_{eff}<-1/3$, the universe will undergo an accelerated phase of expansion.
In addition, Eqs. (\ref{x})-(\ref{l}) were used by the authors to plot the evolution diagram in the $x-y$ parameter space.  However, we find that it is more physical if we use Eqs. (\ref{omega})-(\ref{lambda}) to plot the evolution diagram in the $\gamma_{\phi}-\Omega_{\phi}$ parameter space. In the following section, a comparison between these two approaches will be given.

\section{The comparison of two analyses}\label{comparison}

There are totally five critical points in Table \ref{crit}. Point \textit{O} stands for a matter dominated universe ($\Omega_{m}=1$), but it is a saddle point for all values of $\lambda$. Points\textit{$A_{\pm}$} correspond to the universes dominated by the scalar field kinetic energy with no acceleration, and they are not the stable points. Points \textit{B} and \textit{C} are the more interesting critical points. And then, we will discuss their characteristics in detail in phase diagrams.

\subsection{Solution $B$}

\begin{figure}
\centering
\includegraphics[width=2.5truein,height=2truein]{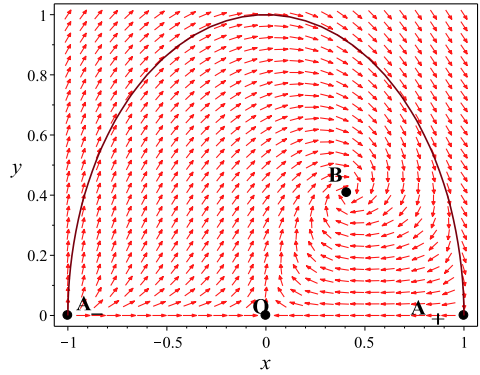}
\caption{Phase space of $x-y$ with $\lambda=3$ and $w_b=0$. Point $B$ is the global attractor. The red line is the unit circle which stands for $\Omega_{\phi}=1$ ($x^2+y^2=1$).} \label{fxy3}
\end{figure}

\begin{figure}
\centering
\includegraphics[width=2.5truein,height=2truein]{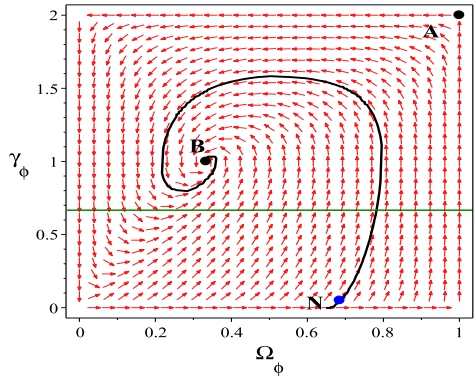}
\caption{Phase space of $\Omega_{\phi}-\gamma_{\phi}$ with $\lambda=3$ and $w_b=0$. The blue point \textit{N} stands for the current state of universe, and the black line is its evolution trajectory with $\ln(a)$ from $-7$ to $10$. The green horizontal line represents $\gamma_{\phi}=2/3$. The region below it stands for accelerated expansion of universe.} \label{f3}
\end{figure}

In the case of point \textit{B}, the scalar field energy density is $0<\Omega_{\phi}=3(1+w_b)/\lambda^2<1$, implying the universe evolves under both the matter and scalar field. This solution is the famous scaling solution where the effective EoS matches the background EoS ($w_{eff}=w_b$). In Fig. \ref{fxy3}, point \textit{B} is plotted in the $x-y$ parameter space, with $\lambda=3$ and $w_b=0$ (dust). Based on the result, point \textit{B} is shown to be a global attractor, since the system eventually evolves to point \textit{B} regardless of the initial conditions. In addition, solution $A_{-}$ (or $A_+$ when $\lambda$ is negative) represents the early universe as the arrows originate from it. However, the physical significance of $x-y$ plane is not enough transparent to describe current universe. Conventionally, solution point \textit{B} is not considered as the representative of current universe, because the EoS of this point, $w_b$, cannot lead to an accelerating expansion. In order to alleviate this problem, the double exponential potential has been considered in Refs. \cite{barreiro2000quintessence,Tsujikawa:2013fta}. But we will see that this solution can actually provide the possibility of cosmic acceleration.

On the contrast, a phase diagram plotted in $\Omega_{\phi}-\gamma_{\phi}$ space can provide more physical information. Taking Fig. \ref{f3} as an example, the phase space analysis is performed and point \textit{B} is shown to be the global attractor still, consistent with the result of Fig. \ref{fxy3}. More importantly, the evolution of the universe could be presented in the $\Omega_{\phi}-\gamma_{\phi}$ phase space due to the cosmological context of the phase diagram. According to the latest observation  \cite{ade2014planck}, it favors the current state of universe with $\Omega_{\phi 0}=0.685, \gamma_{\phi 0}=0.05$ , which is labeled as the blue point \textit{N} (Fig. \ref{f3}). Accordingly, the evolution curve of the universe is obtained using $\Omega_{\phi 0}$ and $\gamma_{\phi 0}$ as the initial conditions. The phase space analysis in $\Omega_{\phi}- \gamma_{\phi}$ plane indicates that the universe would evolve along the black line before reaching the stable point \textit{B}. As mentioned before, the cosmic acceleration requires $w_{eff}<-1/3$, and hence $\gamma_{\phi}<2/3$. As shown in Fig. \ref{f3}, Point \textit{N} evolving to point \textit{B} along with black line will across the green horizontal line, that means the cosmic accelerating expansion will slow down until decelerating forever (state \textit{B}). That is to say this solution could describe the current accelerated expansion of the universe, merely it will slow down until decelerating. This phenomenon is so-called transient acceleration, which is also suggested by SNIa data \cite{2009PhRvD..80j1301S}, and investigated in several theoretical scenarios \cite{qi2016transient,Fabris:2009mn,2008PhRvD..77l3512B,2011CQGra..28l5026G,2013IJTP..tmp..453C}.

Very interestingly, the solutions of non-accelerating expansion are still found useful in presenting the current acceleration of the universe. This is because such solutions describe the final state of the evolution, leaving the intermediate evolutionary state of the accelerating expansion rather untouched. Therefore, the stable point \textit{B} is still valuable, since to trace back current cosmic accelerating expansion is now possible.


On the other hand, the past evolution of the universe should be checked. The evolution with respect to $\ln(a)$ from $-7$ to $10$ corresponding to redshift $z$ from $1095$ to about $-1$ is calculated, which means the history of the universe from early epoch to the future should be tested. Starting from point \textit{N}, the solid black line is found to be discontinuous in the past (Fig. \ref{f3}), implying a singularity in the process of time reversal. But it does not mean that the solution is invalid, and its reason will be discussed in the following Sec. \ref{singularity}.

\subsection{Solution $C$}

\begin{figure}
\centering
\includegraphics[width=2.5truein,height=2truein]{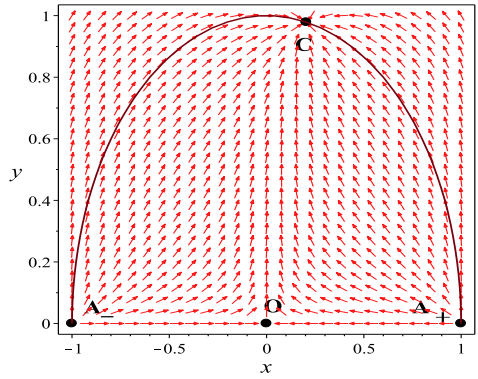}
\caption{Same as Fig. \ref{fxy3} but for $\lambda$=0.5.} \label{fxy0.5}
\end{figure}

\begin{figure}
\centering
\includegraphics[width=2.5truein,height=2truein]{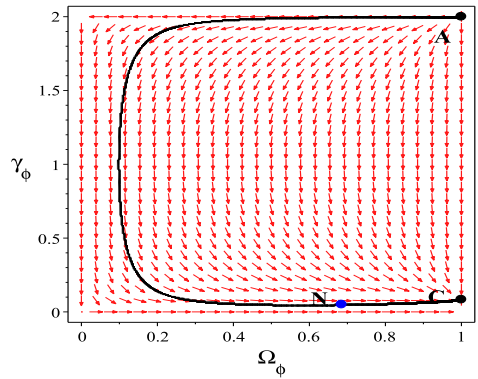}
\caption{Same as Fig. \ref{f3} but for $\lambda$=0.5.} \label{f0.5}
\end{figure}

In Table \ref{crit}, point \textit{C} stands for the universe completely dominated by the scalar field ($\Omega_{\phi}=1$). If only matter ($w_b=0$) is considered, the effective EoS parameter is $w_{eff}=w_{\phi}=\lambda^2/3-1$. For $\lambda^2<2$, the exponential potential scalar field could drive the cosmological acceleration. In Fig. \ref{fxy0.5}, the phase space analysis is performed for $\lambda=0.5$ and $w_b=0$ in $x-y$ parameter space. Point \textit{C} is found to be an attractor. The universe evolves from the initial state $A_-$ or $A_+$ to the final state \textit{C}.

Fig. \ref{f0.5} displays the phase space of universe in $\Omega_{\phi}-\gamma_{\phi}$ parameter space with $\lambda=0.5$. Point \textit{C} is now found to be the global attractor. Because of its dynamic behavior, solution \textit{C} is insensitive to the initial conditions so that it could alleviate the coincidence problem. The universe evolves from decelerating to accelerating expansion. The current state, point \textit{N} with $\Omega_{\phi 0}=0.685$ and $\gamma_{\phi 0}=0.05$, evolves to the global attractor \textit{C}. The EoS parameter $\gamma_{\phi}$ changes slightly, and the universe is expected to accelerate forever, which is like the cosmological constant. In the process of time reversal, point \textit{N} could evolve to the unstable point \textit{A} with no singularity.


\section{Discussion of singularity} \label{singularity}

So far, the advantages of analysis in $\Omega_{\phi}-\gamma_{\phi}$ space have been demonstrated. Nonetheless, the singularity problem just emerged in $\Omega_{\phi}-\gamma_{\phi}$ space, which is abnormal because these two analyses describe the same system. The reason is that, in some cases, the Eqs. (\ref{omega})-(\ref{lambda}) are no longer valid in the evolution of the system. As mentioned previously, the Eqs. (\ref{omega})-(\ref{lambda}) are established in $x\geq0$. In Fig. \ref{fxy3}, any point in $x\geq 0$ evolving to the past will enter into $x<0$ region which is not defined in $\Omega_{\phi}-\gamma_{\phi}$ space, and then there will be a singularity in Fig. \ref{f3}. For Fig. \ref{fxy0.5}, the majority of point in $x\geq 0$ evolving to the past will not enter into $x<0$ region, and then it does not have singularity in Fig. \ref{f0.5}. This indicates that the Eqs. (\ref{omega})-(\ref{lambda}) can't perfectly describe the dynamical system and they have a range of applications.
However they have been used effectively in many studies, therefore, it should be necessary to determine their scope of application. Through the comparison of these two analyses, it shows that the value of $\lambda$ is crucial to the scope. We use numerical method to determine the appropriate value of $\lambda$.

\begin{figure}
\centering
\includegraphics[width=2.5truein,height=2truein]{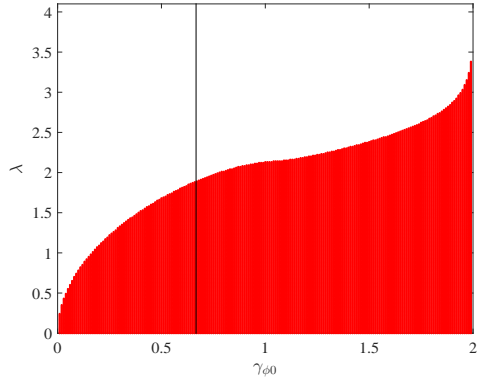}
\caption{The initial conditions $\gamma_{\phi 0}$ and $\lambda$. The shaded red regions show their valid values. The perpendicular line stands for $\gamma_{\phi 0}=2/3$.}\label{fsrl}
\end{figure}

To solve the differential Eqs. (\ref{omega})-(\ref{lambda}) of the system, the initial conditions $\Omega_{\phi 0}$, $\gamma_{\phi 0}$ and $\lambda$ are required. The $\gamma_{\phi 0}-\lambda$ plane is scattered sufficiently for $\Omega_{\phi 0}=0.685$. With these scattered points, the differential Eqs. (\ref{omega})-(\ref{lambda}) could be solved in principle. Whether the effective $\Omega_{\phi} (z)$ and $\gamma_{\phi} (z)$ have singularities will be used to check the validity of the initial conditions $\gamma_{\phi 0}$ and $\lambda$. The result is shown in Fig. \ref{fsrl} with a shaded red region. The perpendicular line is $\gamma_{\phi 0}=2/3$, and the region to the left could lead to an accelerating expansion.
As what could be seen, if the Eqs. (\ref{omega})-(\ref{lambda}) to be used, the initial value of $\lambda$ should be small, because bigger values of $\lambda$ usually lead to singularities in the cosmic evolution back to the past. Form the other perspective, if one moment of the past is chosen as initial point, the cosmological parameters evolving to the present will not pass the region in Fig. \ref{fsrl}.

\begin{figure}
\centering
\includegraphics[width=2.5truein,height=2truein]{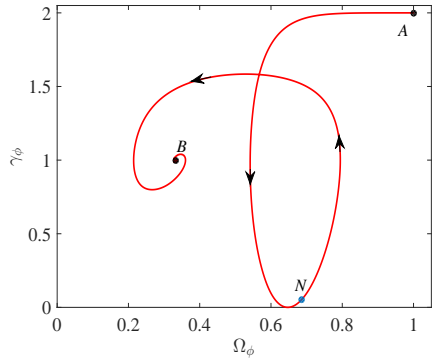}
\caption{The evolution of solution \textit{B} considering $x<0$.}\label{fl32}
\end{figure}

From this perspective, the evolution of solution \textit{B} described by $\Omega_{\phi}-\gamma_{\phi}$ appearing singularity does not mean this solution is invalid. It just has no definition in the past, but it can describe the future. If we consider the situation of $x<0$, the evolution of solution \textit{B} is presented in Fig. \ref{fl32}.

\section{Conclusions and discussions} \label{end}
Till now, the cosmological parameters' space is employed, and the quintessence cosmology with an exponential potential $V(\phi)=V_0\exp(-\lambda\phi)$ is employed to illustrate the perspective. Previously, the scalar fields are analyzed using dynamical system techniques in the $x-y$ parameter space. Now an analysis using a more physically significant parameter space $\Omega_{\phi}-\gamma_{\phi}$ is made possible. By comparing these two analyses, the advantages of this cosmological parameters' space are demonstrated. The information of system solutions could be visually described by its coordinate in the $\Omega_{\phi}-\gamma_{\phi}$ phase space. Given observational data, more information could be obtained. The solution B is the so-called scaling solution. Conventionally, this solution is not supposed to give the cosmic accelerating expansion, but we prove it could give a transient acceleration and this phenomenon has been supported by some observation data. In other words, the stable attractor solutions of the system are not the description of current state of the universe but the final state of cosmic evolution, while the intermediate evolutionary state remained uncertain. Unfortunately, there has been a singularity in evolution of solution \textit{B}. Its reason is that the transformation from $x-y$ to $\Omega_{\phi}-\gamma_{\phi}$ is incomplete, and then a numerical method is used to calculate the range of application.

There are scaling solutions when $\lambda^2>3\gamma_b$ in which the transient accelerating universe is given. While $\lambda^2<3\gamma_b$, the universe is expected to accelerate forever,
which is like the cosmological constant. In either case, the quintessence cosmology could describe the cosmic accelerating expansion in the present, moreover, it could alleviate the coincidence problem for its dynamic behavior. This demonstrates that the quintessence scalar field is a excellent scenario. In addition, a previous strong observational constraint implies  $\lambda\geq 9$ \cite{bean2001early}. Therefore, the transient accelerating universe is more expected.

\section*{Acknowledgments}
We quite appreciate the anonymous referee for his/her suggestions to improve this manuscript. J.-Z. Qi would like to express his gratitude towards Prof. Rong-Jia Yang and Prof. Yun-Gui Gong for their generous help. This work is supported by the National Natural Science Foundation of China (Grant Nos. 11235003, 11175019, and 11178007).

\bibliographystyle{spphys}
\bibliography{Notes}
\end{document}